\documentclass[prb,aps,twocolumn,amsmath,amssymb,floatfix,
superscriptaddress]{revtex4}
\usepackage{amsmath}
\usepackage{amsfonts}
\usepackage{amssymb}
\usepackage[dvips]{graphics}
\usepackage{color}
\usepackage{soul}
\usepackage[colorlinks=true, citecolor=blue, urlcolor=blue ]{hyperref}
\date{\today}

\begin{document}
	
	\title {Uncertainty Relations for the Relativistic Jackiw-Nair Anyon:\\ A First Principles Derivation}
	
	\author{Joydeep Majhi}
	
	\affiliation{Physics and Appli ed Mathematics Unit, Indian Statistical Institute, 203 Barrackpore Trunk Road, Kolkata-700 108, India}
	
	\author{Subir Ghosh }
	
	\affiliation{Physics and Appli ed Mathematics Unit, Indian Statistical Institute, 203 Barrackpore Trunk Road, Kolkata-700 108, India}

	\begin{abstract}
	 joydeepmjh@gmail.com (Joydeep Majhi),  subirghosh20@gmail.com (Subir Ghosh)\\
	 \vskip .5cm
	    In this paper we have explicitly computed the $position-position$ and $position-momentum$ (Heisenberg) Uncertainty Relations for the model of relativistic  particles with arbitrary spin, proposed by  Jackiw and  Nair \cite{jn} as a model for Anyon, in a purely quantum mechanical framework. This supports (via  Schwarz inequality)  the conjecture that anyons  live in a 2-dimensional {\it{noncommutative}} space. We have  computed the non-trivial uncertainty relation between anyon coordinates,  ${\sqrt{\Delta x^2\Delta y^2}}=\hbar\bar{\Theta}_{xy}$, using the recently constructed anyon wave function  \cite{jan}, in  the framework of   \cite{bel}. We also compute the Heisenberg (position-momentum) uncertainty relation for anyons. Lastly we show that the identical {\it{formalism}} when applied to  electrons, yield  a trivial position uncertainty relation,  consistent with their  living  in a 3-dimensional  commutative space.
	\end{abstract}
	\maketitle

	\textbf{Introduction:} 	Consider $x,y$ to be the spatial coordinates for excitations in a $2+1$-dimensional field theoretic model, proposed by Jackiw and Nair \cite{jn} (see also \cite{p248}), that describes relativistic particles of arbitrary spin, purportedly referred to as Anyons \cite{lm,wan1,wan2}, (although its arbitrary  statistics was not shown).   In this paper we	 explicitly show, in a  quantum mechanical framework, that the position uncertainty relation  ${\sqrt{\Delta x^2\Delta y^2}}=\bar{\Theta}_{xy} $  for these anyons  has  a non-zero lower bound for $\bar{\Theta}_{xy}$, where    ${\sqrt{\Delta A^2}}$ is the uncertainty  (dispersion or fluctuation from the mean value i.e. standard deviation) of a generic hermitian operator  $A$. Our result agrees with the commonly used assumption that anyons   live in {\it{Non-Commutative}} (NC) space. 	 The  Schwarz inequality 
	\begin{equation}\label{ineq}
		{\sqrt{\Delta A^2~\Delta B^2}}\geq \frac{1}{2}<[A,B]>,
	\end{equation}
	if considered for $x,y$, indicates $<[x,y]>\neq 0$. An interesting observation is that $\bar{\Theta}_{xy} $ is the minimum effective spatial area occupied by an anyon. We tentatively claim  the $ (length)^2$-dimensional  NC-parameter 	 $\bar{\Theta}_{xy}$ to be  a new physical constant. We also compute the Heisenberg Uncertainty Relation (HUR) for anyon
	${\sqrt{\Delta r^2\Delta p^2}}= \hbar$ where 	$r={\sqrt{x^2+y^2}},~p={\sqrt{p_x^2+p_y^2}}$.  These results for anyons are new.
	
	In deriving the above, we have used a technique, recently formulated   by  Bialynicki-Birula  and  Bialynicka-Birula  \cite{bel}, who compute HUR for  free electrons, using  spinor wave functions (see also \cite{bph1,bph2, bbos}). We follow the same procedure to derive the above uncertainty relations for anyons, making use of the recently constructed   anyon wavefunction  from a work involving the present authors \cite{jan}.
 
Since our demonstration of the counter intuitive non-zero result  $\bar{\Theta}_{xy}$ for anyons is new, we need to show that the result is not a spurious artifact of our formalism. Thus we use the same formalism \cite{bel} to calculate spatial uncertainty relation for conventional $3+1$-dimensional electrons; this turns out to be zero (as expected) thus establishing robustness of the formalism.
 
It is to be noted that noncommutativity of $x,y$ and non-zero $\bar{\Theta}_{xy}$ are two completely different issues and here we are directly concerned  only with the latter.  We stress that possible noncommutative nature of the $x-y$-plane of anyon does not play any role in our study. In fact our result of non-zero $\bar{\Theta}_{xy}$ is (possibly a necessary but)  not a sufficient condition condition for noncommutative  $x-y$; proving noncommutativity is beyond the scope of our work.  It needs to be duly emphasized that the present analysis is done purely in a quantum mechanical framework using anyon wave functions in calculating expectation values $<A>$ because so far, in modelling anyons, an NC algebra (or non-canonical algebra) between anyon dynamical variables were simply posited in a semi-classical formalism. In some variants a generalized (spinning) point particle model was constructed which had constraints that yielded   Dirac brackets,  to be  identified with the NC algebra (see for example \cite{spinan,ghosh,duval,nair,chou,ghosh2,c11}). The NC algebra was geared to generate the arbitrary anyon spin. A non-relativistic limit of the Jackiw-Nair anyon, as suggested in \cite{jn2}, also yields a non-commutative spatial algebra, as proved in \cite{h595}.  In the present work we have provided a totally quantum mechanical treatment for the anyon, based on the Jackiw-Nair anyon model.

	The impact of anyons in theoretical and applied physics is easily established from its ubiquitous role  in High Energy  (via Chern-Simons theory \cite{aa01,aa02}) to Condensed Matter (quantum Hall effect \cite{aa1}, high $T_c$ superconductivity \cite{aa2}) to the exciting arena of non-Abelian anyons (in fault tolerant quantum computation \cite{aa3,aa4}). The posssibility of noncommutative space and its effect on modern physics can be seen in \cite{ncrev,szabo,ban}. Thus a thorough understanding of anyon theory and its living in a noncommutative space is necessary.

	The  quantum mechanical Jackiw-Nair model of   \cite{jn} was extended  in \cite{jan} to the full construction of anyon wave function. The scheme of \cite{bel} is well-suited for our purpose since the Jackiw-Nair equation for anyon is structurally similar to the Dirac equation for electron, both being first order in derivatives. Crucial difference in solutions is that anyon wave function is an infinite component one  although constraints are imposed to reduce it to a single polarization \cite{jn,jan}.	Apart from the anyonic position UR we also compute;\\ (i) ${\sqrt{\Delta r^2\Delta p^2}}=\hbar$ HUR for anyons and (ii) show in the framework of \cite{bel} that $\Delta x^2\Delta y^2=0$ for electron in $3+1$-dimensions  which is indeed reassuring.

	{\bf Outline of the formalism:} 
	
	The probability densities in position and momentum space, for a generic system, are defined as  $\rho_r(\boldsymbol{r})=|\psi(\boldsymbol{r})|^2=\sum_{\beta}|\psi_{\beta}(\boldsymbol{r})|^2 ,~ \rho_p(\boldsymbol{p})= |\tilde \psi(\boldsymbol{p})|^2=\sum_{\beta}|\tilde \psi_{\beta}(\boldsymbol{p})|^2$, where $\tilde \psi (p)$ is the Fourier transform of $\psi (r)$ and $\psi_{\beta}$ are the components of the column vector $\psi$. Now $\psi_\beta (\boldsymbol{r},t)$ can be expressed as
		\begin{equation}
		\psi_\beta (\boldsymbol{r},t)=\int \frac{d^2p}{2\pi}~ F_\beta(\boldsymbol{p})v_\beta(\boldsymbol{p})e^{i(\boldsymbol{p}.\boldsymbol{r}-iE_pt)}
		\label{psi}
		\end{equation}
		where $F_\beta(\boldsymbol{p})$ is orthonormalized free particle solution and $v_\beta(\boldsymbol{p})$ is an arbitrary function with $\beta$ referring to components of the column vector.

	The  standard expressions for the dispersions  $\Delta r^2$ and $\Delta p^2$ in terms of the probability densities in position and momentum space 
	\begin{equation}
		\Delta r^2 = \frac{1}{N^2}\int d^3r (\vec{r}- \langle \boldsymbol{r} \rangle )^2 \rho_r(\boldsymbol{r}) \label{r}
	\end{equation}
	\begin{equation}
		\Delta p^2 = \frac{1}{N^2}\int d^3p (\vec{p}- \langle \boldsymbol{p} \rangle )^2 \rho_p(\boldsymbol{p}) \label{p}
	\end{equation}
	are calculated, 
	where $N^2$ is the normalization constant,
	\begin{equation}
		N^2 = \int d^3r~ \rho_r(\boldsymbol{r}) = \int d^3 p~ \rho_p(\boldsymbol{p}).
	\end{equation}
	Uncertainty relations are always formulated at a fixed time. Without loss of generality, $<\boldsymbol{r}>,~<\boldsymbol{p}>$ can be dropped from (\ref{r},\ref{p}) \cite{bel}. In terms of the wavefunctions, we find
	\begin{equation}
		N^2 = \int d^3p \mid v(\boldsymbol{p}) \mid^2 
	\end{equation}
 where the function $v(\boldsymbol{p})$ of momentum variable represents the independent degrees of freedom of an anyon moving in free space.
 
	$\Delta p^2$ is trivially given by
	\begin{equation}
		\Delta p^2 = \frac{1}{N^2} \int d^3p ~\boldsymbol{p}^2  \mid v(\boldsymbol{p}) \mid^2
	\end{equation}
	whereas  $\Delta r^2$  can be found by using the following identity \cite{bel},
	\begin{equation}
		\boldsymbol{r} \psi_{\alpha}(\boldsymbol{r}) = \int \frac{d^3p}{(2 \pi)^{3/2}}i\boldsymbol{\nabla}_p (F_\alpha v(\boldsymbol{p})) e^{i\boldsymbol{p}\cdot\boldsymbol{r}},
	\end{equation}
	using (\ref{psi}) with $v(\boldsymbol{p})$ being an arbitrary weight function as mentioned above. 
	
		{\bf Anyon wavefunction:} Let us start with the anyon wave function for an anyon of arbitrary spin  $s=1-\lambda$. The dynamical equation for a spin one particle in $2+1$-dimension) in co-ordinate and momentum space $(i \partial_a = p_a)$ is given by \cite{jn}
	
	\begin{equation}
		\partial_a \epsilon^{abc}F_c \pm mF^b = 0 ; ~~~
		(\boldsymbol{p}\cdot \boldsymbol{j})^a_bF^b + ms F^a = 0
	\end{equation}
 with $\boldsymbol{j}$ being the angular momentum operator and m is the mass term. If $\boldsymbol{J}$ is the total spin contribution to the Lorentz generators $\boldsymbol{K}$, then $\boldsymbol{J}= \boldsymbol{K}+\boldsymbol{j}$.
	
	The solution of the three vector $F^a$, $a=1,2,3$ in the Minkowski metric $\eta_{\mu \nu} = \textrm{diag}(1,-1,-1)$ is given by

	\begin{equation}
		F^a(p) = \frac{m}{\sqrt{2}E} \left[ \begin{bmatrix}
			0 \\
			1 \\
			i
		\end{bmatrix} + \frac{p^x +i p^y}{m(E+m)} \begin{bmatrix}
			E+m \\
			p^x \\
			p^y
		\end{bmatrix} \right].
		\label{anm}
	\end{equation}
		This construction has been extended in an elegant way to the  Jackiw-Nair anyon equation \cite{jn}   to describe an anyon of arbitrary spin  $s=1-\lambda$, whose  momentum space dynamics is given by,
	\begin{equation}
		P.(K+j)_{an \hspace{0.1cm}a'n'} f^{\lambda,a'}_{n'} + ms {f}^{\lambda}_{an}=0,~~(D_af^{\lambda,a})_n =0,
		\label{an1}
	\end{equation}
($D^a_{nn'}=\epsilon^a_{~bc}P^bK^c_{nn'}$) where the second equation is the subsidiary (constraint) relation and $n$ runs from 0 to $\infty$. 
	For $\lambda =0$ the anyon reduces to spin one model discussed earlier. The notation and other details can be found in \cite{jan} (see also  \cite{jn}). $K^a$ are the Lorentz generators with action defined as
	\begin{equation}
\begin{aligned}
K^0 {f}^{\lambda +}_n F^a &=(\lambda+n){f}^{\lambda +}_n F^a \\
K^{+}{f}^{\lambda +}_n F^a &=\sqrt{(2 \lambda+n)(n+1)}  {f}^{\lambda +}_{n+1} F^a\\
K^{-} {f}^{\lambda +}_n F^a &=\sqrt{(2 \lambda+n-1) n} {f}^{\lambda +}_{n-1} F^a
\end{aligned}
\end{equation} 
where "$+$" superscript denotes representations bounded below. There is an analogous bounded above representation and  $ K^{\pm}=K^x \mp i K^y$ \cite{jn}. Thus  explicit form of  the free anyon solution, ($\infty\geq n \geq 0$) is given by \cite{jan} 
\begin{eqnarray}
		f_n^{\lambda,a+} =&& \left( \frac{2m}{E+m} \right)^{\lambda} \sqrt{\frac{\Gamma(2\lambda + n)}{n! \Gamma(2\lambda)}} \left( \frac{p^x+ip^y}{E+m}\right)^n \nonumber \\
		&&\times F^a(p)e^{-ipx}
		\label{ann1}
\end{eqnarray}
where $F^a(p)$ is same as the spin-$1$ case defined in (\ref{anm}).
		
Before proceeding further let us ensure: (i) The probability distribution   thus generated is conserved; (ii) The Lorentz generators defined earlier are self-adjoint. Complications can stem from the sum over index $n$ running from $zero$ to $infinity$ and possible appearance of null or negative norm states. These issues are addressed in \cite{val} in detail. Furthermore, in \cite{val} an alternative and more compact anyon model was proposed where, the spin one base used here \cite{jn,jan}, was replaced by a Majorana-Dirac spin $1/2$ base.   \\

(i) Anyon current conservation: We will explicitly derive the conservation law for probability current $\partial^\mu j_\mu^{(s=1-\lambda)}=0$, where $j_0^s$ denotes the probability density (computational steps are provided in Supplemental Material of \cite{jan}). Using explicit form of Anyon wave function (\ref{ann1}) a long calculation yields \cite{jan},
\begin{eqnarray}
&&\partial^0 \sum_{n=0}^{\infty}\left[\left(f_n^{0 \dagger} f_n^0+f_n^{x \dagger} f_n^x+f_n^{y \dagger} f_n^y\right)\right. \nonumber \\
&&-i\left(f_n^{y \dagger} K_{n n^{\prime}}^0 f_{n^{\prime}}^x-f_n^{x \dagger} K_{n n^{\prime}}^0 f_{n^{\prime}}^y\right)-\partial^x \sum_{n=0}^{\infty}\left[\left(f_n^{0 \dagger} f_n^x+f_n^{x \dagger} f_n^0\right) \right. \nonumber \\
&& \left. -i\left(f_n^{y \dagger} K_{n n^{\prime}}^x f_{n^{\prime}}^x-f_n^{x \dagger} K_{n n^{\prime}}^x f_{n^{\prime}}^y\right)\right] -\partial^y \sum_{n=0}^{\infty}\left[\left(f_n^{0 \dagger} f_n^y+f_n^{y \dagger} f_n^0\right) \right.\nonumber \\
&& \left. - i\left(f_n^{y \dagger} K_{n n^{\prime}}^y f_{n^{\prime}}^x-f_n^{x \dagger} K_{n n^{\prime}}^y f_{n^{\prime}}^y\right)\right] =0
\label{acur}
\end{eqnarray}
where the sum over $n$ has been carried out in \cite{jan}. \\

Self-adjointness of Lorentz generators $K^a$:
We have to check whether the following matrix elements are equal  holds for arbitrary $f$ for a specific value of $\lambda$,
\begin{eqnarray}
\left\langle f^\lambda (p_\mu'), K^a f^\lambda (p_\mu\right\rangle=\left\langle K^a f^\lambda (p_\mu ), f^\lambda (p_\mu)\right\rangle .
\end{eqnarray}
Since $K^a$ acts only on the index $n$, the non-trivial part of the matrix element is written following the convention,
\begin{equation}
\begin{aligned}
K_{n n^{\prime}}^0&=\left\langle\lambda, n\left|K^0\right| \lambda, n^{\prime}\right\rangle =(\lambda+n) \delta_{n n^{\prime}}\\
 K_{n n^{\prime}}^{+}&=\left\langle\lambda, n\left|K^{+}\right| \lambda, n^{\prime}\right\rangle \\ &=\left\langle\lambda, n\left|\sqrt{\left(2 \lambda+n^{\prime}\right)\left(n^{\prime}+1\right)}\right| \lambda, n^{\prime}+1\right\rangle \\
&=\sqrt{(2 \lambda+n-1) n} \delta_{n, n^{\prime}+1} \quad \\
 K_{n n^{\prime}}^{-} &=\left\langle\lambda, n\left|K^{-}\right| \lambda, n^{\prime}\right\rangle \\ &=\left\langle\lambda, n\left|\sqrt{\left(2 \lambda+n^{\prime}-1\right) n^{\prime}}\right| \lambda, n^{\prime}-1\right\rangle \\
&=\sqrt{(2 \lambda+n)(n+1)} \delta_{n, n^{\prime}-1} \quad
\end{aligned}
\end{equation} 
We recover 
$$ K^x=\frac{1}{2}\left(K^{+}+K^{-}\right);~~ K^y=\frac{i}{2}\left(K^{+}-K^{-}\right)$$
and obtain in a straightforward way
\begin{subequations}
\begin{align}
\therefore \quad K^0_{nn'}=&(\lambda+n) \delta_{n n^{\prime}} \\
K^x_{nn'}=&\frac{1}{2}\left[\sqrt{(2 \lambda+n-1) n} \delta_{n, n^{\prime}+1}\right. \nonumber \\
&\left. +\sqrt{(2 \lambda+n)(n+1)} \delta_{n, n^{\prime}-1}\right] \\
K^y_{nn'}=&\frac{i}{2}\left[\sqrt{(2 \lambda+n-1) n} \delta_{n, n^{\prime}+1}\right. \nonumber \\
&\left. -\sqrt{(2 \lambda+n)(n+1)} \delta_{n, n^{\prime}-1}\right]
\end{align}
\end{subequations} 
The volume integral in the matrix element introduces $\delta (p'_\mu -p_\mu)$. To calculate $\left\langle f^\lambda (p_\mu'), K^a f^\lambda (p_\mu\right\rangle$, once again we invoke the $n$-summation identities \cite{jan} and get, for $K^0$
\begin{equation}
\begin{aligned}
&\sum_{n,n'=0}^{\infty}\left(f_n^{a+}\right)^{\dagger} K^0_{nn'} f_{n^{\prime}}^{a+}\\
&=\sum_{n=0}^{\infty}(\lambda+n)\left(f_n^{a+}\right)^{\dagger}\left(f_n^{a+}\right) =\frac{E \lambda}{m}
\end{aligned}
\end{equation}
for $K^x$
\begin{equation}
\begin{aligned}
&\sum_{n=0}^{\infty}\left(f_n^{a+}\right)^{\dagger} K^x f_{n^{\prime}}^{a+}=\sum_{n=0}^{\infty}\left[\left(f_n^{a+}\right)^{\dagger} \frac{1}{2} \sqrt{(2 \lambda+n-1) n} f_{n-1}^{a+} \right. \\
& +\left. \left(f_n^{a+}\right)^{\dagger} \frac{1}{2} \sqrt{(2 \lambda+n)(n+1)}\left(f_{n+1}^{a+}\right)\right] =\frac{p_x \lambda}{m}\\
\end{aligned}
\end{equation}
and for  $K^y$
\begin{equation}
\begin{aligned}
&\sum_{n=0}^{\infty}\left(f_n^{a+}\right)^{\dagger} K^y f_{n^{\prime}}^{a+}\\
&=\sum_{n=0}^{\infty}\left[\left(f_n^{a+}\right)^{\dagger} \frac{i}{2} \sqrt{(2 \lambda+n-1) n} f_{n-1}^{a+}\right.\\
&-\left. \left(f_n^{a+}\right)^{\dagger} \frac{i}{2} \sqrt{(2 \lambda+n)(n+1)}\left(f_{n+1}^{a+}\right)\right] =\frac{P_y \lambda}{m} .
\end{aligned}
\end{equation}
In exactly similar manner we calculate $\left\langle K^a f^\lambda (p_\mu ), f^\lambda (p_\mu)\right\rangle$; 
\begin{equation}
\begin{aligned}
\sum_{n,n'=0}^{\infty}\left(K^0_{nn'}f_n^{a+}\right)^{\dagger} k^0 f_{n^{\prime}}^{a+} =\sum_{n=0}^{\infty}(\lambda+n)\left(f_n^{a+}\right)^{\dagger}\left(f_n^{a+}\right)=\frac{E \lambda}{m}
\end{aligned}
\end{equation}
\begin{equation}
\begin{aligned}
&\left(K^x_{nn'} f_n^{a+}\right)^{\dagger}  f_{n^{\prime}}^{a+}\\
=&\sum_{n^{\prime}=0}^{\infty}\left[\left(f_{n^{\prime}-1}^{a+}\right)^{\dagger} \left(\frac{1}{2}\right) \sqrt{(2 \lambda+n^{\prime}-1) n^{\prime}} f_{n^{\prime}}^{a+}\right. \\
& \left. +\left(f_{n^{\prime}+1}^{a+}\right)^{\dagger} \frac{1}{2} \sqrt{(2 \lambda+n^{\prime})(n^{\prime}+1)}\left(f_{n^{\prime}}^{a+}\right)\right]=\frac{P_x \lambda}{m}
\end{aligned}
\end{equation}
\begin{equation}
\begin{aligned}
&\left(K^{y}_{nn'} f_n^{a+}\right)^{\dagger}  f_{n^{\prime}}^{a+}\\
=&\sum_{n^{\prime}=0}^{\infty}\left[\left(f_{n^{\prime}-1}^{a+}\right)^{\dagger} \left(-\frac{i}{2}\right) \sqrt{(2 \lambda+n^{\prime}-1) n^{\prime}} f_{n^{\prime}}^{a+}\right. \\ 
& \left. +\left(f_{n^{\prime}+1}^{a+}\right)^{\dagger} \frac{i}{2} \sqrt{(2 \lambda+n^{\prime})(n^{\prime}+1)}\left(f_{n^{\prime}}^{a+}\right)\right]=\frac{P_y \lambda}{m}
\end{aligned}
\end{equation}
We arrive at identical results as the ones derived above in (18,19,20). This proves the self-adjointness of the Lorentz generators $K^a$.
	
{\bf Spatial uncertainty relation for anyon:} To settle this critical issue, let us consider $\Delta x^2\Delta y^2=\hbar^2 \bar{\Theta}_{xy} ^2$. Computations with anyon wave functions (\ref{ann1}) yields (in polar coordinates) 
\begin{eqnarray}
		&&\Delta x^2= \frac{1}{N^2} \int_0^{\infty}p~dp \int_0^{2\pi} d\theta  \nonumber \\
		&&  \left[ \mid\cos\theta \partial_p v(p,\theta)-\frac{1}{p}\sin\theta\partial_{\theta}v(p,\theta)\mid^2+ \right. \nonumber \\
		&&\frac{1}{2}\left(\frac{\cos^2\theta(m^2+\lambda E^2)}{E^4}+\frac{\sin^2\theta}{p^2}\left(3+4\lambda(\lambda-2) \right. \right. \nonumber \\
		&& \left. \left. +4(\lambda-1)\left(\frac{m}{E}-\frac{\lambda E}{m}\right)+\frac{\lambda(1+2\lambda)p^2}{m^2}\right) \right)  \nonumber \\
		&& \times \mid v(p,\theta)\mid^2 +\frac{i m (\lambda-1)(m-E)+\lambda p^2}{m E p}\sin\theta  \nonumber \\
		&& \left. \times v^*(p,\theta)\left(\cos\theta \overleftrightarrow{\partial_p} -\frac{1}{p}\sin\theta\overleftrightarrow{\partial_{\theta}}\right)v(p,\theta)\right]
	\end{eqnarray}

	Similar results are obtained for $\Delta y^2$, as given in Supplemental Material (III). Following \cite{bel}	we simplify the system by invoking spherical symmetry $v(p,\theta) =v(p)$, to obtain
		\begin{eqnarray}
		\Delta x^2  &&=   \frac{\pi}{N^2}\int_0^\infty dp~p~ \left[|\partial_pv(p)|^2 \right. \nonumber \\
		&&  + \frac{1}{2}\left(\frac{m^2+\lambda E^2}{E^4} \right. +\frac{1}{p^2}\left( 3+4\lambda (\lambda -2) +4(\lambda -1) \right. \nonumber \\
		&& \left. \left. \left. \times \left(\frac{m}{E}-\frac{\lambda E}{m}\right) +\frac{\lambda (1+2\lambda)p^2}{m^2} \right) \right) \mid v(\boldsymbol{p})\mid^2 \right] .
		\label{n1}
	\end{eqnarray}
 One can check that $\Delta y^2$ is also given by the above relation, a consequence of spherical symmetry.
	Note that  we are interested only in the minimum value for $\bar{\Theta}_{xy}$. Hence it is justified to restrict to $v(p)$ only \cite{bel} since angular contributions can only increase $\bar{\Theta}_{xy}$. 
	
	Our aim is to compute $\Delta x^2 \Delta y^2$ and show that it has a non-zero minimum. It needs to be emphasized that only as far as numerical numbers are involved, spherical symmetry dictates that $\Delta x^2  = \Delta y^2$ whereas in reality $\Delta x^2$ and $ \Delta y^2$ are two distinct and independent entities, each pertaining to two coordinate directions that are mutually orthogonal.  Hence we define  $\Delta x^2 \Delta y^2=\hbar^2 \bar{\Theta}_{xy}^2$ keeping in mind that there are two independent parameters involved and the naive equality between $\Delta x^2$ and $\bar{\Theta}_{xy}$ is not to be considered. We obtain a Schrodinger-like  variational equation for $v(\boldsymbol{p})$ whose minimum eigenvalue will provide the cherished value of $\bar{\Theta}_{xy}^2$ \cite{bel};
	\begin{eqnarray}
		\left[-\partial_q^2 -\frac{1}{q}\partial_q +U^{(x,y)}_{(d,\lambda)}(q)\right]v_{(d,\lambda)}(q) &&=\frac{\bar{\Theta}_{xy}}{2} v_{(d,\lambda )}(q) \nonumber \\
        && = \Theta_{xy} v_{(d,\lambda )}(q)~~~
	\label{scc}
	\end{eqnarray}
	where, at this stage we are allowed to use the numerical equality, $\bar{\Theta}_{xy}/(1+\frac{\Delta y^2}{\Delta x^2})=\bar{\Theta}_{xy}/2={\Theta} _{xy}$ 
	and the potential is
	\begin{eqnarray}
		&&2U^{(x,y)}_{(d,\lambda)}(q) =\frac{(1+\lambda(1+d^2q^2))d^2}{(1+d^2q^2)^2}+\frac{1}{q^2}\{3+4\lambda(\lambda -2) \nonumber \\
		&& ~~~ +4(\lambda -1)\frac{1-\lambda(1+d^2q^2))}{{\sqrt{1+d^2q^2}}}+\lambda(1+2\lambda)d^2q^2\}
		\label{ns1}
	\end{eqnarray}
Rescaled variables used are  
$q=\frac{p}{mcd},~  d=\frac{1}{m c}\left(\frac{\hbar^2 \Delta x^2}{\Delta y^2}\right)^{1 / 4}$ \cite{ib}.
Following   \cite{bel} one can consider a non-relativistic limit $m\rightarrow \infty$ or $c\rightarrow \infty$ leading to  $d \rightarrow 0$ (keeping $\Theta_{xy}$ fixed) when there is a surprising cancellation of $\lambda$ terms yielding  $U^{(x,y)}_{(0,0 )}(q)=-1/(2q^2)$,
	\begin{eqnarray}
		\left[-\partial_q^2 -\frac{1}{q}\partial_q -\frac{1}{2q^2}\right]v_0(q) =\Theta_{xy(0)} v_{0}(q).
		\label{1scc}
	\end{eqnarray}

\begin{figure}[ht]
		{\centering \resizebox*{8.cm}{8.7cm}{\includegraphics{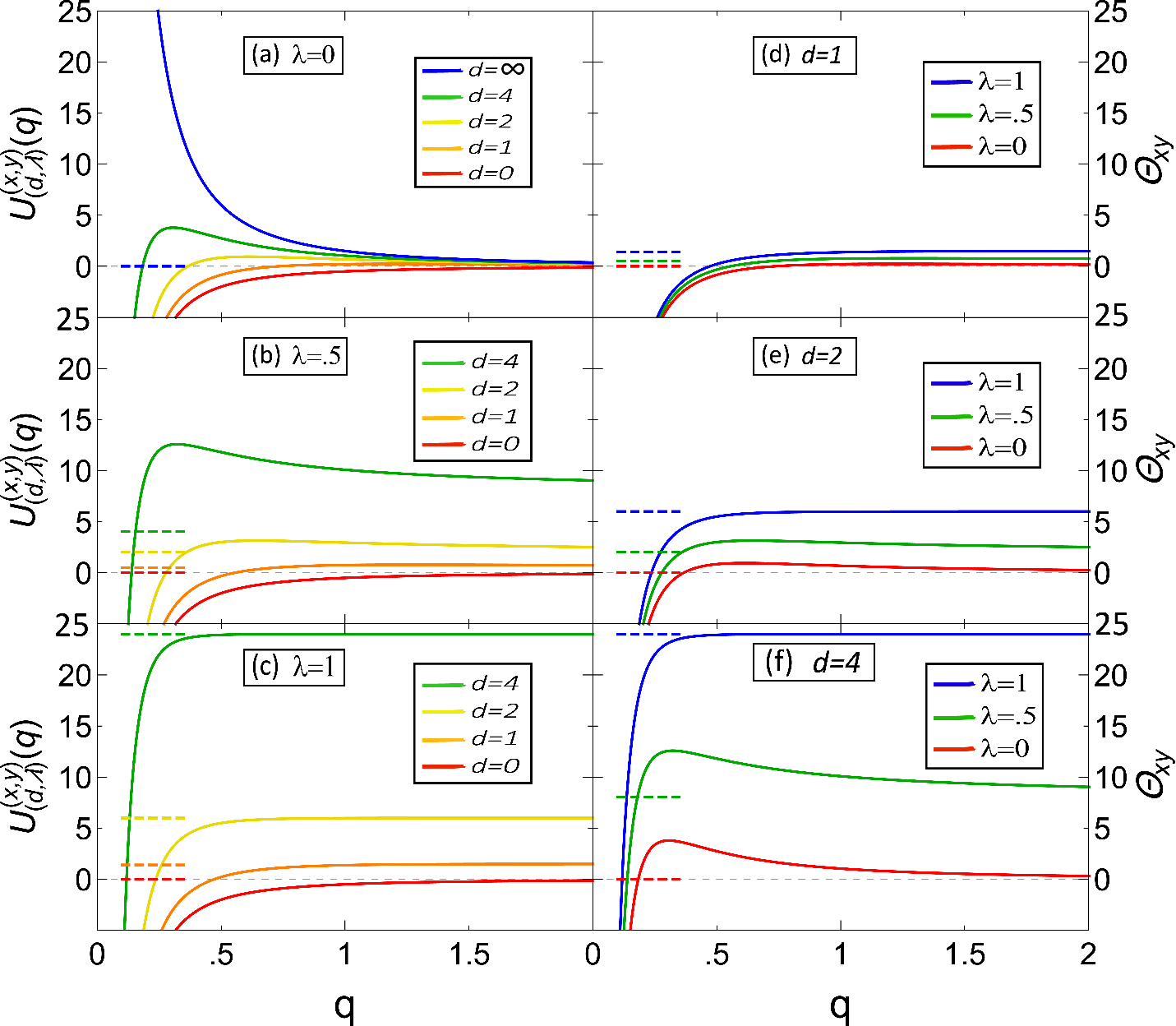}}\par}
		\caption{Variation of potential $U^{(x,y)}_{(d,\lambda )}(q)$ with $q$ for different values of $d$ and $\lambda$. Heights of the dashed short horizontal lines in each figure indicate the lowest eigenvalues for $\Theta_{xy}$  for the corresponding (color matched) potential. }
		\label{fig5}
	\end{figure}

\begin{figure}[ht]
	{\centering \resizebox*{7.9cm}{6.1cm}{\includegraphics{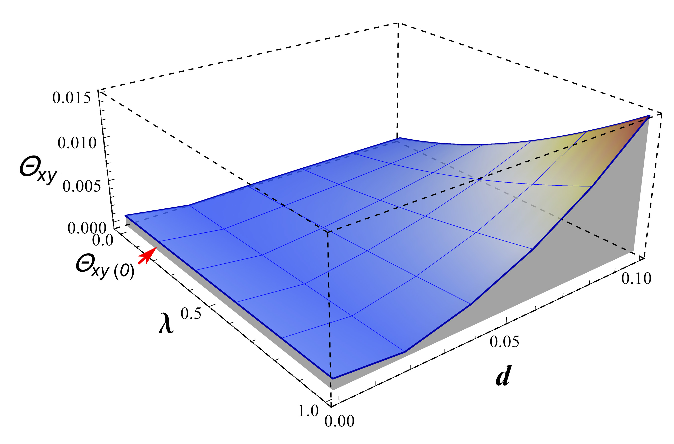}}\par}
	\caption{ $\Theta-d-\lambda$ diagram shows a non-zero minimum value   $\Theta_{xy(0)}\approx 0.002$ .}
	\label{fig3}
\end{figure}

	We are only interested in finding the lowest positive eigenvalue $\Theta_{xy(0)} $. The dependence of
	the potential profile  on $d$ and $\lambda$ are shown in  in Fig.(\ref{fig5}). Few of the lowermost eigenvalues of $\Theta_{xy}$ are shown in each of the panels. The numerically computed eigenvalues for $\Theta_{xy}$ are revealed in the $\Theta_{xy}-d-\lambda$ diagram in 
	the lowest positive eigenvalues on the parameters $d$ and $\lambda$ is shown in Fig.(\ref{fig3}). Notice that the minimum value $\Theta_{xy}\approx 0.002~(time/mass)$, the dimension comes from the definition $\Delta x^2\Delta y^2=\hbar^2\bar{\Theta}^2_{xy}$ and also agrees with the dimension coming from (\ref{1scc}) and above, $dimension ~|1/q^2|=dimension~ \bar{\Theta}_{xy}$ . As explained above, this corresponds to  $d=0$  which can be contrasted to the commutative space for electron discussion, (see below (\ref{nc1})). This is our most significant result showing that, thanks to the Schwarz inequality, the claim that {\it{anyons  live in a noncommutative space}} is mathematically consistent.
	
	{\bf Heisenberg Uncertainty Relation for anyon:} The details of the computation of $\Delta r^2$ are provided in Supplemental Material. As before, we exploit the spherical symmetry by introducing polar coordinates,  
	\begin{equation}
		N^2 = \int^{\infty}_0 dp~ p\int^{2\pi}_0 d\theta ~ \mid v(p,\theta) \mid^2
		d\theta\end{equation}
	
	\begin{equation}
		\Delta p^2 = \frac{1}{N^2} \int^{\infty}_0 dp~ p\int^{2\pi}_0 d\theta~ p^2\mid v(p,\theta) \mid^2
	\end{equation}
	
	\begin{eqnarray}
		\Delta r^2 &&=  \frac{1}{N^2} \int^{\infty}_0 dp~ p\int^{2\pi}_0 d\theta \nonumber \\
		&& \left[p \mid \partial_p v(p,\theta)\mid^2 + \frac{1}{p}\mid \partial_{\theta}v(p,\theta)\mid^2 \right. \nonumber \\
		&& + \left(-\frac{p^3}{E^4}+\lambda ~p \left(\frac{1}{E^2}+\frac{2\lambda + 1}{m^2}\right) \right. \nonumber \\
		&& \left. -\frac{4(\lambda-1)(E-m)(E\lambda+m)}{mEp}\right)\mid v(p,\theta)\mid^2 \nonumber \\
		&& - i\frac{m(\lambda-1)(m-E)+\lambda p^2}{mEp}v^*(p,\theta)\overleftrightarrow{\partial_{\theta}}v(p,\theta)].
	\end{eqnarray}
	Following  \cite{bel} we define $\bar{\Theta}_{xp} = \sqrt{\Delta r^2 \Delta p^2}/\hbar $ and try to find the minimum value of $\bar{\Theta}_{xp}$ by varying $v^*$ and equating it to zero. For the same reasons mentioned earlier we invoke spherical symmetry since angular contributions can only increase the eigen value  \cite{bel}. The variational equation for  $\Delta r^2 \Delta p^2$ with respect to $v^*$ reduces to
	
	\begin{eqnarray}
		&& \left[  \Delta p^2 \left(- \partial^2_p-\frac{1}{p}\partial_p - \frac{p^2}{E^4}-\frac{4(\lambda-1)(E-m)(E\lambda+m)}{Emp^2} \right. \right.   \nonumber \\
		&& \left. \left. +\lambda\left(\frac{1}{E^2}+\frac{2\lambda+1}{m^2}\right)\right)+p^2\Delta r^2-2\bar{\Theta}_{xp}^2\right]v(p)=0 .
	\end{eqnarray} 
	Shifting to dimensionless variables \cite{bel}  $q=\frac{p}{mcd},~ d=\frac{1}{mc}\left(\frac{\hbar^2\Delta p^2}{\Delta r^2}\right)^{1/4}$ we recover a Schrodinger-like equation for $v_{(d,\lambda )}(q)$
	\begin{eqnarray}
		\frac{1}{2}\left[-\partial_q^2 -\frac{1}{q}\partial_q + 2U^{(r,p)}_{d,\lambda}(q)\right]v_{(d,\lambda )}(q)
		=\bar{\Theta}_{xp} v_{(d,\lambda )}(q)~~~~~
		\label{sc}
	\end{eqnarray}
	where the potential is
	\begin{eqnarray}
		&&2U^{(r,p)}_{(d,\lambda )}(q) = -\frac{q^2d^4}{(1+q^2d^2)^2}+\lambda d^2\left(\frac{1}{q^2d^2+1}+2\lambda+1\right) \nonumber \\
		&& -\frac{4(\lambda-1)(\sqrt{q^2d^2+1}-1)(\lambda\sqrt{q^2d^2+1}+1)}{q^2\sqrt{q^2d^2+1}}+q^2
	\end{eqnarray}

	\begin{figure}[ht]
	{\centering \resizebox*{8.cm}{6.cm}{\includegraphics{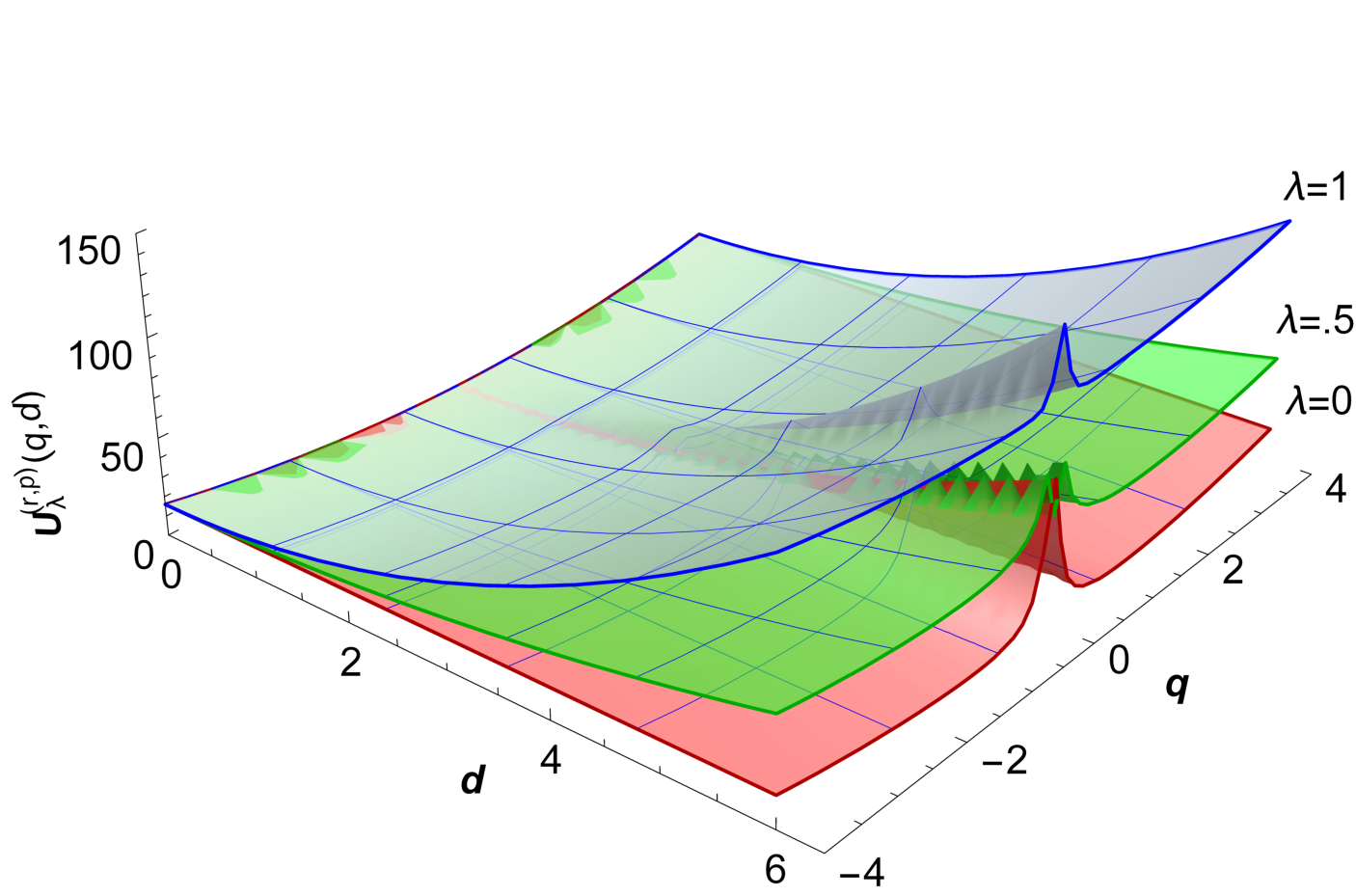}}\par}
	\caption{Variation of $U^{(r,p)}_{(d,\lambda)}(q)$ with $q$ and $d$ for different values of $\lambda$ .}
	\label{fig2}
\end{figure}
	
	\begin{figure}[ht]
		{\centering \resizebox*{8.1cm}{8.7cm}{\includegraphics{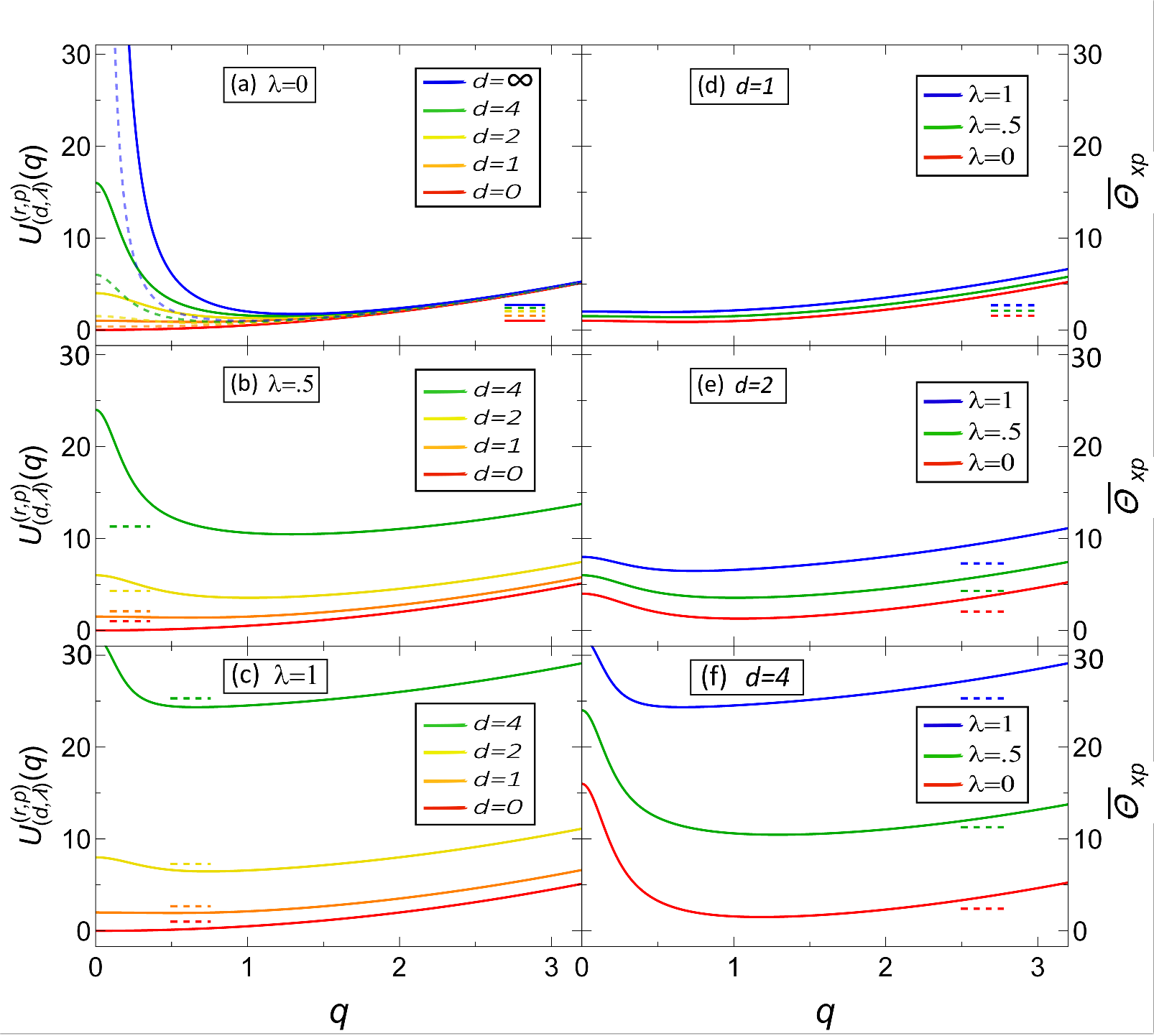}}\par}
		\caption{Potential $U^{(r,p)}_{(d,\lambda)}$ vs. $q$ is plotted for different values of $d$ and $\lambda$. Heights of the dashed short horizontal lines in each figure indicate the lowest eigenvalues for $\bar{\Theta}_{xp}$  for the corresponding (color matched) potential. In Fig. (a), the eigenvalues for the limiting cases $d=0$ and $d=\infty$ for $\lambda=0$ are denoted by solid lines. Dotted curves in (a) represent the  potential term for electron, as given in Ref.\cite{bel}, for the same values of $d$.}
		\label{fig1}
	\end{figure}
	Notice that $d\rightarrow 0$ yields the non-relativistic limit, for which (\ref{sc}) becomes independent of $\lambda $, 
	\begin{eqnarray}
		\frac{1}{2}\left[-\partial_q^2 -\frac{1}{q}\partial_q +q^2\right] v_0(q) = \bar{\Theta}_{xp(0)} v_0(q)
		\label{vo}
	\end{eqnarray}
	Incidentally (\ref{vo}) is identical to the corresponding equation for electron in \cite{bel} apart from a factor of $2$ in the $\partial_q/q$ term for space dimensional mismatch. 
	The solution of the above equation is a Gaussian 
	$v_0(q) = exp(-q^2/2)$ with the lowest eigenvalue $\bar{\Theta}_{xp(0)} = 1$. {\it{We recover the HUR for anyon as}} $ \sqrt{\Delta r^2 \Delta p^2}=\hbar $, due to the two (spatial) dimensional nature of the system. This is one of our major results.
	
	It is interesting to note that the relativistic limit   $d \rightarrow \infty$  exists  only for $\lambda =0$ (i.e. for spin $s=1-\lambda =1$, otherwise some $\lambda$-dependent terms diverge) leading to
	
	\begin{eqnarray}
		&&\frac{1}{2}\left[-\partial_q^2 -\frac{1}{q}\partial_q +q^2 + \frac{3}{q^2}\right] v_{(\infty,0)}(q) = \bar{\Theta}_{xp(\infty,0)} v_{(\infty,0)}(q) \nonumber \\
		~&&
	\end{eqnarray}
	The  solution of the above differential equation is 
	$v_{(\infty ,0)}(q) = q^{\sqrt{3}}e^{-q^2/2}$
	and $\bar{\Theta}_{xp(\infty ,0)}=1+\sqrt{3}$. A few representative results are given in tabular form  
	 	\begin{table}[ht]
	 	\begin{tabular}{|c|c|c|c| } 
	 		\hline
	 		~ & $\lambda = 0$ & $\lambda = 0.5$ & $\lambda = 1.0$ \\
	 		\hline
	 		d=0 & 1.0 & 1.0 & 1.0 \\ 
	 		d=1 & 1.54 & 2.08 & 2.69 \\ 
	 		d=2 & 2.05 & 4.29 & 7.29 \\
	 		d=4 & 2.41 & 11.29 & 25.31 \\
	 		d=$\infty$ & 2.73 &  -   &  -  \\
	 		\hline
	 	\end{tabular}
	 	\caption{Lowest eigen values for different values of $d$ and $\lambda$}
	 	\label{tab1}
	 \end{table}
	
	In  Fig.(\ref{fig2})  a  three dimensional plot for $ U^{(rp)}_{(d,\lambda)}(q) - q -d$ for different $\lambda$ shows how the potential separates in to sheets for each $\lambda$. 	 In Fig.(\ref{fig1}) variations of $U^{(rp)}_{(d,\lambda)}$ with $d$ and $\lambda$ are shown along with corresponding eigenvalues $\bar{\Theta}_{xp}$.	The anyon  results are new and  can be compared with similar results for electron \cite{bel}. These constitute  our second major result for anyon.

	{\bf{Anyon equation of state and the NC parameter $\Theta_{xy}$ :}} Using the relations derived above  let us consider
	\begin{equation}
	    \label{i0}
	 \frac{\Theta_{xy}}{\Delta x^2} =\frac{\Theta_{xy} \hbar^2}{\Delta p^2} \approx \frac{\Theta_{xy} \hbar^2}{mkT}  .
	\end{equation}
	One might identify the above with the thermal de Broglie wavelength $\lambda ={\sqrt{(2\pi \hbar^2)/(mkT)}}$. This can play a part in the discussion below.
	
	In statistical mechanics, ideal gases where particles are treated as non-interacting, play an important role. In the quantum version, the Fermi/Bose statics effect is taken in to account in the otherwise non-interacting Fermi/Bose gas. These analysis require the multiparticle wave functions that are simply products of single particle wave functions for Fermions/Bosons. However, following the same procedure for anyons is not possible because single anyon wave functions do not yield the multi-anyon wave function in any simple way. Thus, one considers an ideal anyon gas as an interacting Bose/Fermi gas. The generic equation of state of an ideal gas in two space dimensions, is
$(PA)/(NkT)=ln {\cal{Z}}$. 	$P,A$ are the pressure and area respectively, $N$ the particle number and $T$ the temperature.  ${\cal{Z}}$ is the grand partition function. However, for anyons, considering an interacting system of  Boson/Fermion the equation of motion is expressed as a series
	\begin{equation}
	    \label{i22}
	    \frac{P}{\nu kT}=1+a_2(\nu\lambda^2)+a_3(\nu\lambda^2)^2+~....
	\end{equation}
	where $\nu =N/A$ is the number density, $\lambda$ the thermal wavelength and $a_i$ the (dimensionless) virial coefficients, leading to
	\begin{equation}
	    \label{i2}
	    \frac{P}{\nu kT}=1+a_2(\nu\Theta_{xy})+a_3(\nu \Theta_{xy})^2+~....~.
	\end{equation}
	(For a detailed discussion see \cite{khare}.) The above is a heuristic presentation of a possible significance of $\Theta_{xy}$, the anyonic spatial uncertainty parameter.

	{\bf{Commutative space for electrons:}} Let us check the consistency of our scheme by recovering the commutative space for electrons, in the framework of \cite{bel}. Using free spinor solutions  $F_\alpha^{+}$ for electron, the dispersion for $x$ and $y$ are straightforward to obtain. 	The expression simplifies considerably by invoking spherical symmetry $(f(p,\theta ,\phi )=f(p))$,
	\begin{eqnarray}
		\Delta x^2 =\Delta y^2 &&= \frac{4\pi}{3N^2}\int_0^{\infty} dp~ \sum_s \left[ p^2 \mid\partial_p f(p)\mid^2 \right. \nonumber \\
		&& \left. + \left(1 + \frac{m}{E} + \frac{m^2p^2}{4E^4}\right)\mid f(p)\mid^2 \right]
	\end{eqnarray}
	
	Let us define 
	\begin{equation}
		\Delta x^2\Delta y^2 = \hbar^2\bar{\Theta}_{xy(e)}^2 
		\label{v1}
	\end{equation}
 
	where $\bar{\Theta}_{xy(e)}$ is  the NC parameter of dimension $(length)^2$ (if it turns out to be non-zero) and $e$ stands for electron. Integrating over $\theta$ and $\phi$ and considering the variation with respect to $f^*$ we have 
	
	\begin{eqnarray}
		\left[\Delta x^2 \left( -\partial_p^2-\frac{2}{p}\partial_p +\left(\frac{1}{p^2}-\frac{m}{Ep^2}\right. \right.\right.&&\left.\left.\left.+\frac{m^2}{4E^4}\right)\right) \right]f(p) \nonumber \\
		&& = \Theta_{xy(e)}^2 f(p)
	\end{eqnarray}
	where, $\Theta_{xy(e)}=\frac{\bar{\Theta}_{xy(e)}}{2}$
	Introducing the dimensionless  variable $q = \frac{p}{mcd}$ and parameter $ d=\frac{1}{m c}\left(\frac{\hbar^2 \Delta x^2}{\Delta y^2}\right)^{1 / 4}$, the above equation reduces to
	\begin{eqnarray}
		\left[-\partial_q^2 -\frac{2}{q}\partial_q + V^{(x,y)}_{d}(q)\right]f_{d}(q)
		=\Theta_{xy(e)} f_{d}(q)~~~
		\label{nc1}
	\end{eqnarray}
	\begin{eqnarray}
		V^{(x,y)}_d(q) = \frac{1}{q^2}-\frac{1}{q^2\sqrt{1+d^2q^2}}+\frac{d^2}{4(1+q^2d^2)^2}~~
	\end{eqnarray}
	
	\begin{figure}[ht]
		{\centering \resizebox*{7.5cm}{5.5cm}{\includegraphics{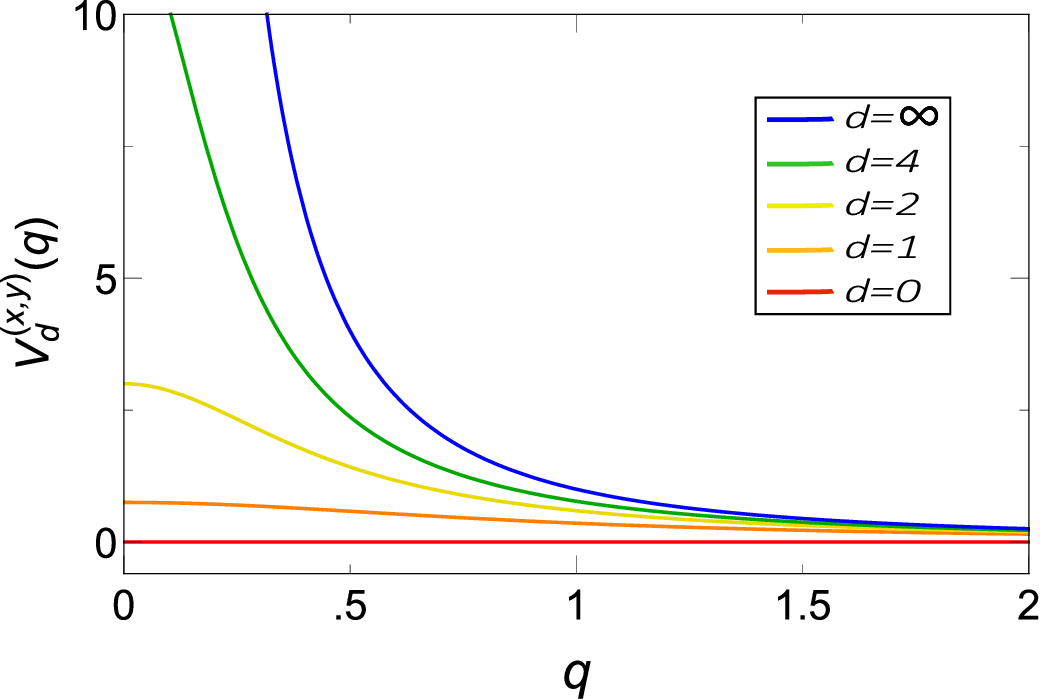}}\par}
		\caption{Variation of the potential $V^{(x,y)}_d(q)$ with $q$ for different values of $d$. }
		\label{fig0_1}
	\end{figure}
	
	In Fig.(\ref{fig0_1}), the profile  of  $q~vs.~V^{(x,y)}_d(q)$  for different values of $d$ is shown. For  $\infty \geq d \geq 0$, the potential approaches zero for large  $q$.
	
	In the two limits of  $d$,   Eq.(\ref{nc1}) reduces to
	\begin{eqnarray}
		&& \left[-\partial_q^2 -\frac{2}{q}\partial_q\right]f_{0}(q)
		=\Theta_{xy(e,0)} f_{0}(q),~~d\rightarrow 0 ~~~~~~~~~~~~~~~~ \\
		&& \left[-\partial_q^2 -\frac{2}{q}\partial_q +\frac{1}{q^2}\right]f_{\infty}(q)
		=\Theta_{(e,\infty)} f_{\infty}(q),~d\rightarrow \infty .
		\label{nc2}
	\end{eqnarray}
	We are interested in the smallest eigenvalue and clearly $\Theta_{xy(e,0)} <\Theta_{xy(e,\infty)} $ (due to the positive contribution from $1/q^2$ in (\ref{nc2}) and the lowest possible value for $\sigma_0$ is zero consistent with free particle solution  $f_0 \sim exp[-i\sqrt{\Theta_{xy(e,0)}}q]$	 for large $q$.   Comparison with electron $\Delta r^2\Delta p^2$ HUR results in \cite{bel} reveals that the presence of $q^2$  (harmonic oscillator) term in the overall potential  that induced a non-zero minimum eigenvalue, is absent here in $\Delta x^2\Delta y^2$ HUR for electron. Thus $3+1-dim$ {\it {electrons live in a commutative space}}. Indeed, this is not a new result but rederived in the present framework, to be contrasted with the anyon result, derived earlier.

	{\bf{Conclusion and future prospects:}} To summarise, we have computed minimum values of products of dispersions between coordinates $x,y$ and between coordinate and momentum $r,p$ for  $2+1$-dimensional arbitrary spin particles, referred to as  anyons, utilising explicit form of anyon wavefunction \cite{jan},   in the framework of \cite{bel}. Non-zero value for the former yields the spatial uncertainty relation for anyon and strongly suggests  that anyons live in noncommutative space. Incidentally we also show, in the same formalism, that Dirac electrons live in commutative space, which is reassuring. We have briefly indicated how $\Theta_{xy}$ might appear in anyon equation of state.
	
	Indeed more research is required but still, it is tempting to interpret the noncommutativity parameter $\bar{\Theta}_{xy}$ as a new and independent constant for planar quantum physics, similar to $\hbar$ in phase space. It might be interesting to consider Zitterbewegung effect for anyons and also introduce  Foldy-Wouthuysen or Newton-Wigner coordinates in the study of anyon. 
	
	More and more applications of anyons in modern physics, (especially in the area of quantum computing), makes it necessary to grasp the  underpinnings of  anyon theory at the microscopic level. We plan to extend our work to model an action principle for anyon that can be generalized to non-abelian anyons, the ultimate goal of this project.
			
	{\bf{Acknowledgements:}} It is indeed a pleasure to thank Professor Iwo Bialynicki-Birula for patiently explaining many subtleties of their work.\\
		{\bf{Data Availability Statement:}} No Data associated in the manuscript.


\begin{thebibliography}{99}    \bibitem{jn} R.Jackiw and V.P.Nair, Phys.Rev.D \textbf{43},1933(1991).
						\bibitem{p248} M.S. Plyushchay, Phys.Lett.B \textbf{248},107(1990).
\bibitem{lm} J.Leinaas and J.Myrheim,
Nuovo Cimento Soc.Ital.Fis.37B,1(1977).
\bibitem{wan1}F.Wilczek, Phys.Rev.Lett. \textbf{49},957(1982).
		\bibitem{wan2} \textit{Fractional Statistics and Anyon Superconductivity}, F.Wilczek, Editor. 
		\bibitem{jan} J.Majhi, S.Ghosh and S.K.Maiti,  Phys.Rev.Lett.\textbf{123}, 164801(2019).  
		\bibitem{bel} I.Bialynicki-Birula and Z.Bialynicka-Birula,
		New J.Phys.\textbf{21},07306(2019).		 			\bibitem{bbos}I.Bialynicki-Birula and A.Prystupiuk,
		Phys.Rev.A\textbf{103},052211(2021).
				\bibitem{bph1} I.Bialynicki-Birula and Z.BialynickaBirula, 
		Phys.Rev.Lett.\textbf{108},140401(2012).
				\bibitem{bph2} I.Bialynicki-Birula and Z.Bialynicka-Birula,
		Phys.Rev.A \textbf{86},022118 (2012).							\bibitem{spinan} M.Chaichian, R.G.Felipe, and D.L.Martinez,
			Phys.Rev.Lett.\textbf{71},3405 (1993);Erratum ibid.\textbf{73},2009(1994). 
			\bibitem{ghosh} S.Ghosh,Phys.Rev.D\textbf{51},5827(1995);Erratum-ibid.D52:4762,1995. 
			\bibitem{duval} C.Duval and P.A. Horvathy,Phys.Lett.B\textbf{479}, 284(2000). 
			\bibitem{nair} V.P.Nair and A.P. Polychronakos, Phys.Lett.B \textbf{505},267 (2001).
			\bibitem{chou} C.Chou, V.P.Nair and A.P.Polychronakos, Phys.Lett.B \textbf{304},105(1993). 
			\bibitem{ghosh2}  S.Ghosh, Phys. Lett.B\textbf{338},235(1994); Erratum-ibid.\textbf{347}468(1995).
				\bibitem{c11} J.L.Cortes and M.S.Plyushchay, Int.J.Mod.Phys. A \textbf{11},3331(1996).
    \bibitem{jn2}R.Jackiw and V.P.Nair, Phys.Lett B551,166(2003).
				\bibitem{h595} P.A.Horvathy and M.S.Plyushchay, Phys.Lett.B \textbf{595},547(2004).
						 \bibitem{aa01}S.Deser, R.Jackiw and S.Templeton, Phys.Rev. Lett.\textbf{48},975 (1982).
		 \bibitem{aa02} A.J.Niemi and G.W.Semenoff, Phys.Rev.Lett. \textbf{51}, 2077(1983).
		\bibitem{aa1} A.Stern, \textit{Anyons and the quantum Hall effect - a pedagogical review}, cond-mat/0711.4697.
		\bibitem{aa2}E.J.ferrer, R.Hurka, V.de la  Incera, 		  Mod.Phys.Lett.B \textbf{11},1(1997). 
				\bibitem{aa3}A.Yu.Kitaev, Ann. Phys.(N.Y.)\textbf{303}, 2 (2003).
		\bibitem{aa4}S. Rao, (2017) Introduction to abelian and non-abelian anyons; In S.Bhattacharjee, M.Bandyopadhyay (eds) Topology and Condensed Matter Physics. Texts and Readings in Physical Sciences, vol 19. Springer, Singapore. 
		\bibitem{ncrev} M.R.Douglas and N.A. Nekrasov, Rev.Mod.Phys.\textbf{73},977 (2001). 
		\bibitem{szabo} R.J.Szabo, Phys.Rep. \textbf{378},207(2003).
		\bibitem{ban} R.Banerjee, B.Chakraborty, S.Ghosh, P.Mukherjee and S.Samanta, Found.Phys.\textbf{39}, 1297(2009).
			\bibitem{val}  P.A.Horváthy, M.S. Plyushchay, M.Valenzuela, Ann.Phys. 325 (2010)1931.
	\bibitem{khare} A.Khare, {\it{Fractional Statistics and Quantum Theory}}, 2005 World Scientific Publishing Co. 
\bibitem{ib} {Professor I.Bialynicki-Birula, private correspondence} 

\end{thebibliography}
\end{document}